\documentclass[%
superscriptaddress,
 prl,
rsi,%
amsmath,amssymb,
reprint,%
]{revtex4-1}

\usepackage{graphicx}
\usepackage{dcolumn}
\usepackage{xcolor}
\usepackage{float}

\begin{document}
\author{Kalyani Chordiya}
\affiliation{ELI-ALPS, ELI-HU Non-Profit Ltd., Wolfgang Sandner utca 3., Szeged, H-6728, Hungary}
\affiliation{Institute of Physics, University of Szeged, Dóm tér 9, H-6720 Szeged, Hungary}
\author{Victor Despr{\'e}}
\affiliation{Theoretische Chemie, PCI, Universit{\"a}t Heidelberg, Im Neuenheimer Feld 229, D-69120 Heidelberg, Germany}
\author{Mousumi U. Kahaly}
\email{Mousumi.UpadhyayKahaly@eli-alps.hu}
\affiliation{ELI-ALPS, ELI-HU Non-Profit Ltd., Wolfgang Sandner utca 3., Szeged, H-6728, Hungary}
\affiliation{Institute of Physics, University of Szeged, Dóm tér 9, H-6720 Szeged, Hungary}
\author{Alexander I. Kuleff}
\email{alexander.kuleff@pci.uni-heidelberg.de}
\affiliation{ELI-ALPS, ELI-HU Non-Profit Ltd., Wolfgang Sandner utca 3., Szeged, H-6728, Hungary}
\affiliation{Theoretische Chemie, PCI, Universit{\"a}t Heidelberg, Im Neuenheimer Feld 229, D-69120 Heidelberg, Germany}

\title{Distinctive Onset of Electron Correlation in Molecular Tautomers}

\date{\today}

\begin{abstract}
We investigate the attosecond response of the electronic cloud of a molecular system to an outer-valence ionization. The time needed for the remaining electrons to respond to a sudden perturbation in the electronic structure of the molecule is a measure of the degree of electron correlation. Using the \textit{ab initio} multielectron wave-packet propagation method, we show that this response time can be sensitive to the molecular structure and the symmetry of the ionized molecular orbital. Our analysis revealed differences in the temporal signatures of ultrafast charge migration processes in the context of tautomeric hydrogen in multinuclear systems, and the relevance of distinctive onset of electron correlation within the overall molecular complexity.
\end{abstract}

\maketitle

Governed by the long-range Coulomb interaction, the motion of the electrons in atoms and molecules is correlated, making that perturbations to a single electron are felt by the whole electronic cloud even in extended systems. The response of the system to the changes introduced by the perturbation often involves rearrangements in its electronic structure. The electron correlation is thus the driving force of a plethora of processes taking place in many-electron systems \cite{Sansone2012CPC}, with autoionization \cite{Autoionization}, population of satellite states upon ionization \cite{cederbaum1986correlation}, energy transfer \cite{Scholes_FRET2003,ICD_Review2020}, and charge migration \cite{CEDERBAUM1999205,Kuleff2018Ultrafast} being just a few examples of such electron-correlation effects. That is why, over the years the electron correlation and the processes driven by it have been a subject of intensive research, both by theory and experiment. So far they have been mainly investigated in the energy domain, where the direct access to the full quantum information (amplitudes and phases) is often difficult. With the advent of the attosecond pulse generation techniques \cite{calegari2016} and the possibility to perform pump-probe experiments with extreme temporal resolution \cite{ramasesha2016}, however, the scientific community obtained a powerful tool to analyze and study electron-correlation processes directly in time. 

One of the important questions in this respect is how fast does the electronic cloud respond to the process-triggering perturbation? Or, in other words, what is the time scale of the electron correlation and how much it is system-specific? In their seminal paper \cite{breidbach2005universal}, Breidbach and Cederbaum studied this question by investigating the response of a many-electron system to a sudden removal of one of its electrons. They showed that the time needed for the electronic cloud to respond to a sudden ionization is about 50~attoseconds (1~as$=10^{-18}$~s) and that this time is nearly independent of the system. Due to the characteristic evolution of the density of the created hole charge, this universal time was interpreted as the time scale of the filling of the exchange-correlation hole of the removed electron. The exchange-correlation hole represents the region of space around each electron where the probability to find another electron is close to zero due to the quantum exchange and correlation effects \cite{parr1989w}. Although the time for filling of the exchange-correlation hole of an initially created vacancy might be universal, it is not the shortest response time of the electronic cloud to a sudden ionization. Later studies showed \cite{Kuleff2007Tracing} that it can be even shorter ($\sim 30$~as), suggesting that this response time nevertheless might depend on the degree of correlation.

\begin{figure*}
\centering
\includegraphics[scale=0.15]{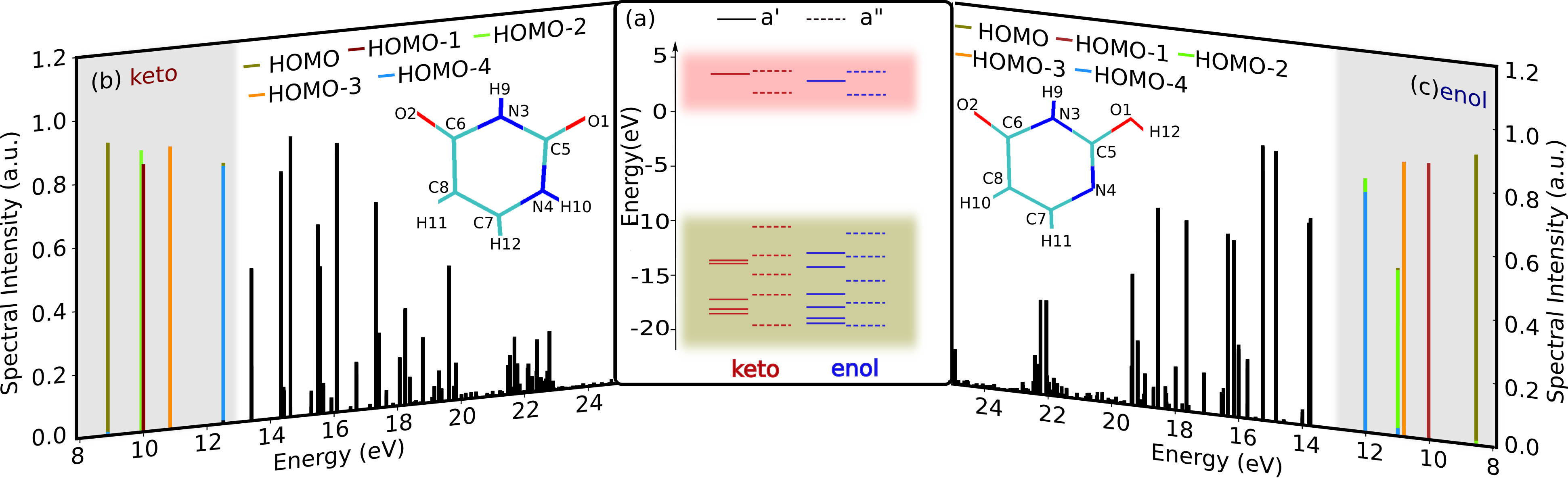}
\caption{(a) Comparison of the energies of the molecular orbital belonging to a' (solid lines) and a" (dashed lines) symmetry of the two tautomers keto (red) and enol (blue) of uracil molecule. Ionization spectra of \textbf{U}-keto (b) and \textbf{U}-enol (c) computed using the Green's function non-Dyson ADC(3) approach. The shaded regions in panel (b) and (c) show the states that will be mostly populated by removing an electron from one of the highest 5 molecular orbitals and investigated in this work.}
\label{fig:Introduction}
\end{figure*}

Here we show that this response time can also be sensitive to the structural differences in the same molecule. For this purpose, we studied the evolution of the charge density of the hole left by a sudden removal of different electrons from two tautomers of the uracil molecule (\textbf{U}). Uracil, a very important component of RNAs, has multiple tautomeric forms \cite{rejnek2005correlated} amongst which the most stable ones (energy difference of 0.48~eV) are `keto' and `enol'. In keto (with two carboxyl groups) the tautomeric hydrogen (H10) bonds with nitrogen (N4), while in enol (with one carboxyl group and one hydroxyl group) the tautomeric hydrogen (H12) bonds with oxygen (O1), see the structures in Fig.~\ref{fig:Introduction} (b,c). The hydrogen, denoted as H10 in keto and H12 in enol (see Fig~\ref{fig:Introduction}) is referred to as tautomeric hydrogen (Ht) throughout this Letter. Both these tautomers are found in C$_s$ symmetry and thus have two types of orbitals -- symmetric (a') and anti-symmetric (a") with respect to the molecular plane.

The two tautomers have similar electronic structures and ionization spectra (see Fig.~\ref{fig:Introduction}). The ionization spectra presented have been computed with the help of the non-Dyson algebraic diagrammatic construction (ADC) scheme \cite{nd-adc3} for approximating the one-particle Green's function. In the ionization spectra, shown in Fig.~\ref{fig:Introduction}(b and c), each line represents a cationic eigenstate with position corresponding to its ionization energy and spectral intensity to its ionization cross-section. In a configuration-interaction picture, each normalized cationic state is represented as a sum of all contributing one-hole (1h) configurations, two-hole--one-particle (2h1p) configurations, etc., with the spectral intensity given by the sum of the weights of all contributing 1h configurations \cite{cederbaum1986correlation}. As the 2h1p and the higher configurations describe excitations on top of the removal of a particular electron, their weight is a measure of the correlation effects contributing to the corresponding state. The first few lines (in the shaded part of Fig.~\ref{fig:Introduction}(b and c)) have large spectral intensities and thus correspond to states in which the correlation effects are relatively small. This is also the regime for which the third-order non-Dyson ADC approach used [ADC(3)] is the most accurate \cite{trofimov2005}. We will, therefore, perform electron-dynamics calculations, allowing to simulate the response of the electronic cloud to the removal of an electron from the highest 5 molecular orbitals (MOs) leading to the population of the cationic states in this region.

In order to trace the response of the system upon ionization, we computed the time-dependent hole density $Q(\vec{r}, t)$ \cite{CEDERBAUM1999205,doi:10.1063/1.1540618} defined as the difference between the electronic density of the system before ionization, $\rho_0 (\vec{r})$, and that after removing an electron from the respective MO, $\rho_i (\vec{r},t)$
\begin{equation}
    Q( \vec{r},t) = \underbrace{\langle\Psi_{0}|\hat{\rho}|\Psi_{0}\rangle}_{\rho_0 (\vec{r})} - \underbrace{\langle\Phi_{i}(t)|\hat{\rho}|\Phi_{i}(t)\rangle}_{\rho_i (\vec{r},t)}.
\label{eq:q(r,t)_final}
\end{equation}
In the above equation, $|\Psi_{0}\rangle$ is the electronic ground-state of the system, and $|\Phi_{i}(t)\rangle$ is the non-stationary state created by suddenly removing an electron out of $i$-th MO. In the Heisenberg picture, the time-dependent electronic density can thus be written as
\begin{equation}
    \rho_i (\vec{r},t) = \langle\Phi_{i}(0)|e^{i\hat{H}t}\hat{\rho}e^{-i\hat{H}t}|\Phi_{i}(0)\rangle,
\label{eq:rho_i}
\end{equation}
where $\hat{H}$ is the cationic Hamiltonian of the system. Using the molecular orbitals of the neutral system as a basis, one can obtain the following expression for the hole density \cite{doi:10.1063/1.1540618}
\begin{equation}\label{eq:holeOcc}
    Q( \vec{r},t) = \sum_{p} |\tilde{\varphi}_{p}( \vec{r},t)|^2 \tilde{n}_{p}(t),
\end{equation}
where $\tilde{\varphi}_{p}( \vec{r},t)$ are the natural charge orbitals and $\tilde{n}_{p}(t)$ their hole-occupation numbers. At each time point, the natural charge orbitals are different expansions in the neutral MO basis set. The present calculations were performed using the multielectronic wave-packet propagation technique \cite{kuleff2005multielectron}. The method uses the non-Dyson ADC(3) scheme to build the cationic Hamiltonian \cite{nd-adc3}, employed then to propagate the initial state via the short iterative Lanczos technique \cite{leforestier1991comparison,park1986unitary}. Further technical and theoretical details about construction and analysis of the hole density can be found in Refs.~\cite{doi:10.1063/1.1540618,kuleff2005multielectron,breidbach2007migration}.

The molecular geometries were optimized at the PBE0/def2-TZVP level of theory, and the Hartree-Fock MOs, orbital energies and two-electron integrals, needed for constructing the ADC Hamiltonian, were generated with Gamess-UK package \cite{guest2005gamess} using cc-pVDZ basis set \cite{dunning1989gaussian}.

\begin{figure*}
\centering
\includegraphics[scale=0.5]{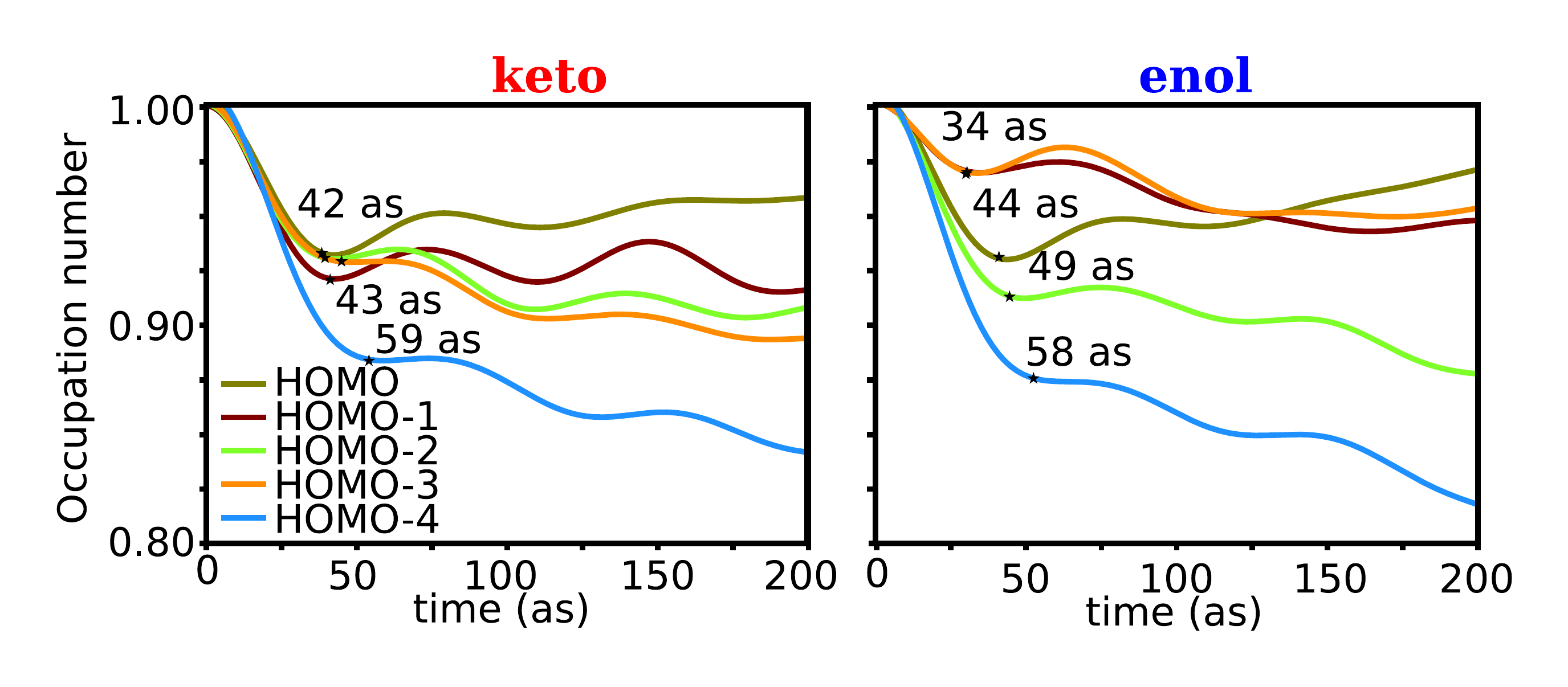}
\caption{Temporal evolution during the first 200~as of the hole-occupation numbers $\tilde{n}_i(t)$, Eq.~(\ref{eq:holeOcc}), after ionization of the five highest occupied molecular orbitals. The corresponding response times are marked with ``*''.}
\label{fig:tempevo}
\end{figure*}

Let us now examine the response of the two uracil tautomers upon ionization out of the five highest occupied MOs. Figure~\ref{fig:tempevo} shows the first 200~as of the evolution of the hole-occupation numbers $\tilde{n}_i(t)$ of the natural charge orbital bearing the initial hole in each of the studied cases, see Eq.~(\ref{eq:holeOcc}). The two a' orbitals show very similar behavior in both keto (HOMO$-2$, HOMO$-3$) and enol (HOMO$-1$, HOMO$-3$). The response time is, however, substantially shorter in enol (34~as) compared to that in keto (42~as). The difference between the tautomers is much less pronounced if we compare the a" orbitals: HOMO, HOMO$-1$, and HOMO$-4$ in keto, and HOMO, HOMO$-2$, and HOMO$-4$ in enol. However, the different a" orbitals show different response times, varying from about 40 to about 60~as. Interestingly, the response time increases when going deeper in the electronic shells, but at the same time larger fraction of the initial hole gets filled during these first instances. We will return to this point below.

Overall, the creation of an a' hole shows a faster response than when an a" electron is removed. A possible explanation of this observation could be that the $\sigma$ orbitals of the molecular skeleton belong to the a' irrep, which facilitates the overlap between the MOs of this symmetry \cite{hoffmann1971interaction} and thus the interaction between the electrons belonging to these orbitals. We can thus expect that in more conjugated systems, the time needed for the electronic cloud to respond to a sudden perturbation will be shorter than in less conjugated molecules. A dependence of the time scale of an electronic process on the degree of electron correlation has been recently observed in the study of non-local electronic decays through carbon chains \cite{Mullenix2020}. It is nevertheless remarkable to see that hints for such a dependence exist already in the immediate response of the electronic cloud to the removal of an electron. 

\begin{figure*}
\centering
\includegraphics[scale=0.13]{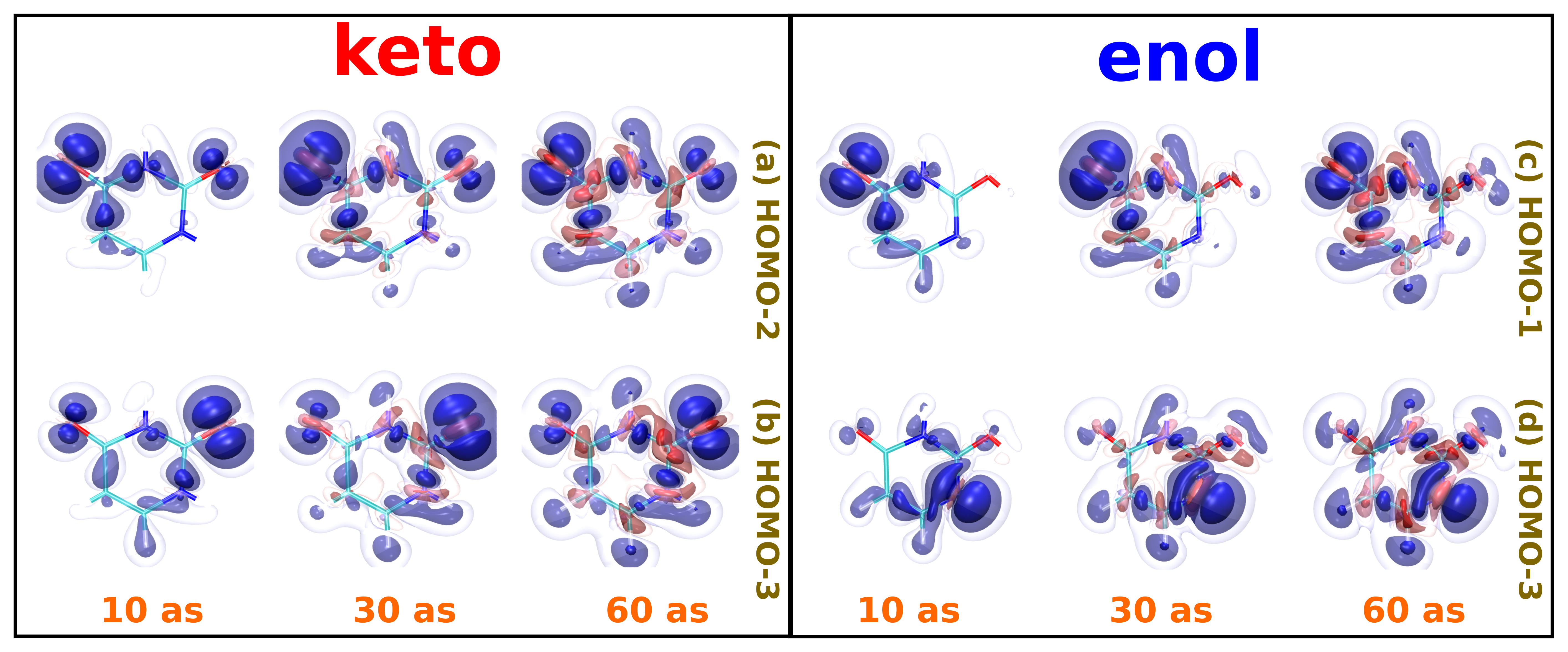}
\caption{Snapshots of the evolution of the hole density, following sudden ionization out of the two outermost molecular orbitals belonging to a' symmetry, (a, b) in keto, and (c, d) in enol at 10 as, 30 as, and 60 as. The hole density is depicted in blue and the electron density, or the regions with excess of electrons, in red.}
\label{fig:sym isosurface}
\end{figure*}

In order to investigate further the differences in the response of the two tautomers upon sudden ionization, it is insightful to analyze the evolution of the hole density and its variation around each atomic site. Snapshots of the evolution of the hole density during the first 60~as following ionization out of the orbitals belonging to a' symmetry are shown in Fig.~\ref{fig:sym isosurface}. We see that the sudden removal of an electron from the highest a' MO leads to a very similar response in the two tautomers, Fig.~\ref{fig:sym isosurface} (a) and (c). Electron density (depicted in red in Fig.~\ref{fig:sym isosurface}) starts to build up mainly around O2 and C6, as well as around C8 and N3, while the hole density increases around H9 and H11(keto)/H10 (enol), and the N3-C5 and C7-C8 bonds, as well as around H12(keto)/H11(enol) and the tautomeric hydrogen (H10 in keto and H12 in enol). The removal of an electron from the next a' MOs triggers different dynamics in the two molecules. The respective orbitals are localized on different sites of the molecules -- in keto the hole is mostly around O1, while in enol it is around N4 (see Fig.~\ref{fig:asym isosurface}(b) and (d)). The response in keto represents a transfer of electron density from C7-H12 to O1-C5 with an excess of electron density forming around N4. In enol, the filling of the exchange-correlation hole is mainly coming from electrons around H9-N3 and O1-H12, the tautomeric hydrogen. Electron density accumulates also around C5 and C7.

The response of the electronic cloud to the sudden removal of the three outermost a" orbitals shows similar tendencies. The snapshots of the respective hole density evolution can be found in the Supplementary Material (SM), Fig.~\ref{fig:asym isosurface}. Electronic density builds up around the atom where the largest fraction of the initial hole has been localized and around the atoms in its immediate proximity. The electron flow is mostly from regions not showing significant hole density initially, confirming that the dynamics mostly represent the filling of the exchange-correlation hole of the initial vacancy. Again, the density variations in keto are somewhat stronger than in enol, and the tautomeric hydrogen is usually an electron donor (except for HOMO) in the process.

In order to better compare the density variations around the different atoms, we computed the rate of these variations. Those are depicted in Figs.~\ref{fig:rates} in the SM. The time needed to reach the maximum rate of density change (the first minimum/maximum in $\frac{\partial Q(\vec{r},t)}{\partial t}$), which we will call maximum-flux time ($\tau$), can be an important measure of the local correlation effects, as it shows the time scale on which a given atomic site starts to donate or receive electronic density. It is intuitively clear that this property is related to the electronegativity of the corresponding atom. Indeed, the maximum-flux time at the oxygens is always somewhat shorter than that at the nitrogens, and the largest density variations are observed at the sites characterized by higher electronegativity, like N and O (see Tab.~\ref{tab:flux time}, where the maximum-flux times at the sites of largest density change are listed). Moreover, the electron density flow is preferentially from the sites with lower electronegativity (carbon and hydrogen) to the sites with a higher one (oxygen and nitrogen).
\begin{table}[h]
    \centering
        \caption{The overall electronic response time to the removal of an electron from the five outermost molecular orbitals of keto and enol, and the maximum flux time ($\tau$) at the atomic site with the highest charge density response given in brackets.}\label{tab:flux time}
    \begin{tabular}{ccccc} \hline
       tautomer & molecular & symmetry & response  & maximum-flux  \\ 
                & orbital   &          & time [as] & time [as] (atom)\\
       \hline\hline
         & HOMO & a" & 42 & 24 (N4) \\ 
         & HOMO$-1$ & a" & 43 & 18 (O1) \\ 
    keto & HOMO$-2$ & a' & 42 & 18 (O2) \\  
         & HOMO$-3$ & a' & 42 & 18 (O1) \\ 
         & HOMO$-4$ & a" & 59 & 18 (O1) \\ 
        \hline
         & HOMO & a" & 44 & 24 (N4) \\ 
         & HOMO$-1$ & a' & 34 & 18 (O2) \\ 
    enol & HOMO$-2$ & a" & 49 & 19 (O1) \\  
         & HOMO$-3$ & a' & 34 & 21 (N4) \\ 
         & HOMO$-4$ & a" & 58 & 17 (O1) \\ 
         \hline
    \end{tabular}
\end{table}
The closer comparison of the rates of change at the respective sites between the two tautomers reveals, however, some differences (see Table.~\ref{tbl: flux a'} and Table.~\ref{tbl: flux a"}). Apart from the ionization out of the HOMO, the electronic cloud around the tautomeric hydrogen always responds differently in the two tautomers. The rate of density variation shows that Ht is an electron donor. The maximum-flux time at Ht is typically reached faster in enol than in keto.

Let us now return to the observed increase in the response time when the hole is created in deeper a" orbitals (see Fig.~\ref{fig:tempevo}). Going deeper into the electronic shells typically increases the correlation effects and thus one would expect the opposite trend, i.e. a decrease of the response time. That the correlation effects increase in the present case can be deduced by the decreasing spectral intensity of the main lines populated by removing an electron from the corresponding a" orbitals (see Fig.~\ref{fig:Introduction}). As we mentioned above, the smaller the spectral intensity of a given state is, the larger the contribution of the 2h1p configurations describing the electron correlation is. The analysis of the rate of variations at the individual sites shows, however, that although the overall partial filling of the initial vacancy proceeds slower, the maximum-flux time is reached faster for the deeper lying orbitals. For both tautomers we obtain that the maximum-flux time changes from 24~as for HOMO to 17~as for HOMO$-4$ (see Tab.~\ref{tab:flux time}). This means that the process of filling of the exchange-correlation hole of the initial vacancy, especially when this vacancy is delocalized over several atomic sites, can be rather sensitive to the local correlation effects at the particular chemical element.  

The calculations discussed above have been performed at a fixed nuclear geometry and although it might be evident that the nuclear motion is too slow to influence the attosecond response of the electronic cloud, it is interesting how much the effect is sensitive to the momentary position of the nuclei at the time of ionization. To investigate this effect we performed calculations at different nuclear geometries around the equilibrium one. Our calculations performed even at maximal deformations in the ground vibrational state along the different vibrational modes show no sensitivity of the response time to these small variations of the atomic positions. The zero-point energy spread of the molecular wave function is, therefore, not expected to smear out the effect.

In summary, our results show that the response of the electronic cloud of a molecular system to a sudden perturbation is not universal and might depend on the strength of the electron correlation, the symmetry for the involved molecular orbitals, and the molecular structure. Although the two studied tautomers of \textbf{U} are structurally and electronically close, the time that the remaining electrons need to respond to a sudden removal of the two outermost $\sigma$-electrons can be quite different ($\sim 42$~as in keto versus $\sim 34$~as in enol). Moreover, a careful analysis of the local variations of the electronic cloud clearly suggest that the process of filling of the exchange-correlation hole can be sensitive to the degree of correlation at a given site and that the flow of electronic density is typically from the atoms with lower electronegativity to the atoms with a higher one. A prominent example of the latter is the response of the tautomeric hydrogen which is a donor of electronic density (except in case of HOMO), but the time and the degree of density variation are different in the different cases studied. All this clearly demonstrates that computing and analyzing the response of a molecule to a sudden removal of an outer-valence electron can be a valuable source of information on the overall strength of the correlation effects in the studied system and their distribution among the system constituents. 

Before concluding, we would like to touch upon the possibility for an experimental investigation of the ultrafast response of the electronic cloud to a sudden perturbation. To be able to study such effects experimentally, we should be able to both ``suddenly'' remove an electron from the system and then measure the following ultrafast charge redistribution with an extreme precision. Although each of these prerequisites can be achieved to some extent, to the best of our knowledge, currently there are no developed techniques that can satisfy both conditions. The sudden-ionization limit can be approached, for example, by ionization with a high-energy photon \cite{ruberti2018}, in which case the ionized electron leaves the interaction region nearly instantaneously. Different interferometric techniques have been used to measure photoionization time delays \cite{schultze2010,klunder2011,eckle2008} that have provided unprecedented precision within several attoseconds. Even zeptosecond accuracy was recently achieved in measuring single-photon two-electron ionization of H$_2$ \cite{Doerner2020zepto}. Nevertheless, a direct measurement of the response of the electronic cloud to a sudden ionization will be very challenging and thus a stringent test for the fast developing attosecond technology.

\section*{Acknowledgements}
ELI-ALPS is supported by the European Union and co-financed by the European Regional Development Fund (GINOP-2.3.6-15-2015-00001). KC and MUK acknowledge PaNOSC European project and Project no. 2019-2.1.13-TÉT-IN-2020-00059 which has been implemented with the support provided from the National Research, Development and Innovation Fund of Hungary, financed under the 2019-2.1.13-TÉT-IN funding scheme. VD and AIK acknowledge financial support from the DFG through the QUTIF priority programme. 

\bibliography{uracil}
\clearpage
\newpage
\Large{\textbf{Supplementary Material}\setcounter{section}{0}} \\
\small
\setcounter{figure}{0}
\setcounter{table}{0}
\renewcommand{\theequation}{S\arabic{equation}}
\renewcommand{\thefigure}{S\arabic{figure}}
\renewcommand{\thetable}{S\arabic{table}}
\vspace{0.3cm}

\begin{figure*}
\centering
\includegraphics[scale=0.13]{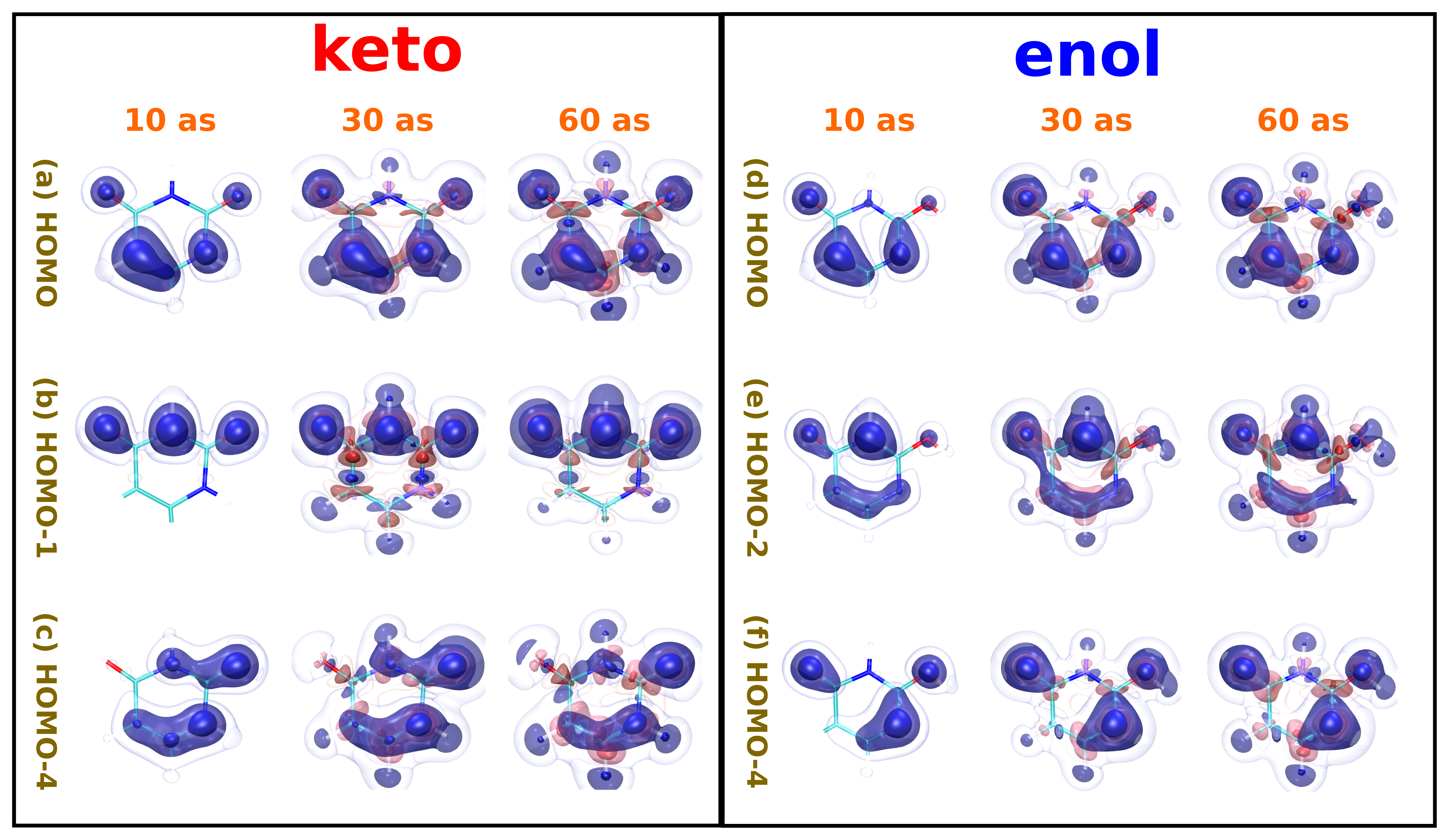}
\caption{Snapshots of the evolution of the hole density, following outer-valence ionization from molecular orbitals belonging to a" symmetry, (a-c) in keto, and (d-f) in enol at 10 as, 30 as, and 60 as. The hole density is depicted in blue and the electron density, or the regions with excess of electrons, in red. 
}
\label{fig:asym isosurface}
\end{figure*}

The response to the ionization out of the three highest a" orbitals is shown in Fig.~\ref{fig:asym isosurface}. In both tautomers, the corresponding MOs are localized either on C7-N4 segment (HOMO), on O1-N3-O2 (HOMO$-1$ in keto and HOMO$-2$ in enol), or on O1-N3 and C7-N4 (HOMO$-4$). Interestingly, the ionization out of these orbitals show rather similar response in both tautomers. We see that the electron density filling the exchange-correlation hole comes mainly from H9 and C7-Ht in all three case, irrespective of the initial hole localization. We observe, however, a difference in the maximum-flux time (see Table.~\ref{tab:flux time} in the main text). After removing an electron from the HOMO (localized on C7-N4), the maximum-flux time is found to be 24~fs in both keto and enol, while it is about 18~fs after removing an electron from one of the lower-lying a" orbitals. These results show that the ultrafast response of the system is actually rather sensitive to the local environment of the removed electron. The maximum-flux time and the charge density variation depend on the degree of electron correlation in the ionized electronic shell, as well as on the symmetry of the involved orbital.

\begin{figure*}
\centering
\includegraphics[scale=0.5]{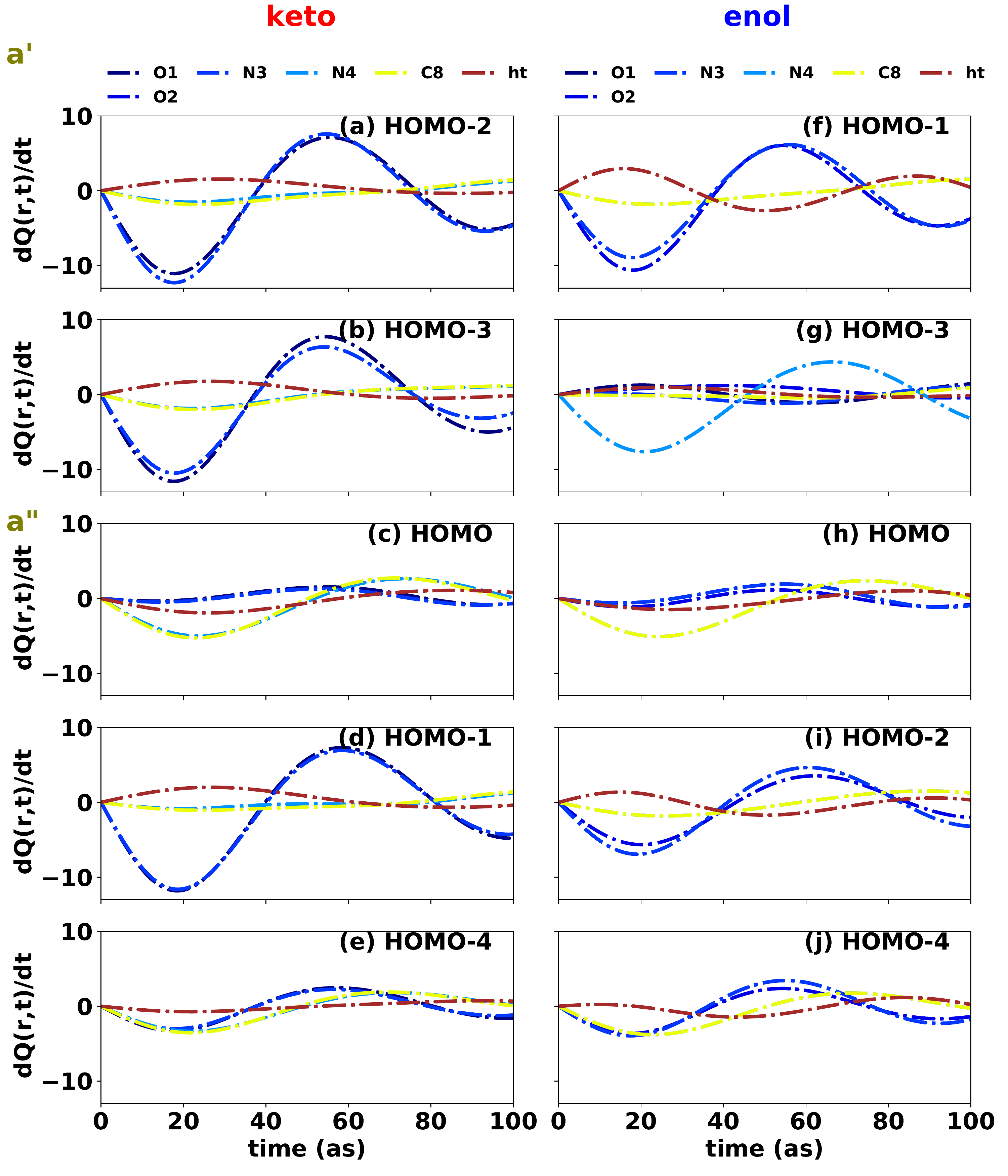}
\caption{Rate of change in hole density, $dQ(\vec{r},t)/dt$, at selected molecular sites. }
\label{fig:rates}
\end{figure*}

To analyze the local effect we summarize the overall rate of change in hole density or flux with respect to time in Fig.~\ref{fig:rates}. Here we have selected the molecular sites showing prominent flux. Additionally, corresponding maximum-flux time is tabulated for a' orbital in Table.~\ref{tbl: flux a'} and for a" in Table.~\ref{tbl: flux a"}. Note, that the molecular site C8 is selected so that the only neighbouring atoms to it are carbon and hydrogen.  

\clearpage
\newpage

\begin{table*}
\centering
\begin{tabular}{|c|c|c|c|c|c|c|}
\hline
          a' & atoms &              dQ(r,t)/dt & time (as)& a'&               dQ(r,t)/dt & time (as) \\
\hline

keto & & && enol &&\\
\hline
   HOMO-2 &    O1 &     -11.07 &        18 & HOMO-1  &  -8.94 &        18 \\
   &    O2 &     -12.28 &        18 & & -10.62 &        18 \\
   &    N3 &     -12.28 &        18 &  &    -8.94 &        18 \\
   &    N4 &      -1.53 &        22 &   &   -1.81 &        23 \\
   &    C8 &      -1.82 &        23 &     & -1.81 &        23 \\
   &    Ht &       1.56 &        29 &      & 2.96 &        16 \\
   \hline
 HOMO-3  &    O1 &     -11.59 &        18 & HOMO-3 &       1.26 &        20 \\
   &    O2 &     -10.49 &        18 &   &     1.2 &        41 \\
   &    N3 &     -10.49 &        18 &&     -   &    -     \\
   &    N4 &      -1.85 &        22 &&      -7.63 &        21 \\
   &    C8 &      -1.98 &        23 &&         - &        - \\
   &    Ht &       1.78 &        26 & &        0.98 &        26 \\
 \hline
\end{tabular}
\caption{Tabulated time (as) at which maximum flux (dQ(r,t)/dt) is observed at different atomic sites for keto and enol a' orbitals. No values are given when no substantial charge flux is present for the given atomic site.}
\label{tbl: flux a'}
\end{table*}

\begin{table*}
\centering
\begin{tabular}{|c|c|c|c|c|c|c|}
\hline
          a" & atoms &              dQ(r,t)/dt & time (as)& a"&               dQ(r,t)/dt & time (as) \\
\hline

keto & & && enol & & \\
 \hline
  HOMO   &    O1 &      - &        - &HOMO&       -0.6 &        15 \\
     &    O2 &      - &        -&&      -1.12 &        17 \\
     &    N3 &      - &        -&&       - &        - \\
     &    N4 &      -5.01 &        24&&      -5.09 &        24 \\
     &    C8 &      -5.26 &        23&&      -5.09 &        24 \\
     &    Ht &       -1.93 &        26&&       -1.48 &        27 \\
\hline
 HOMO-1  &    O1 &     -11.82 &        18& HOMO-2&      -6.93 &        19 \\
   &    O2 &     -11.63 &        19& &      -5.65 &        20 \\
   &    N3 &     -11.63 &        19&&      -6.93 &        19 \\
   &    N4 &      - &        -&&      -1.81 &        25 \\
   &    C8 &      -1.06 &        24&&      -1.81 &        25 \\
   &    Ht &       2.02 &        27&&        1.37 &        15 \\
\hline
   HOMO-4  &    O1 &      -3.26 &        18&HOMO-4&      -3.94 &        17 \\ 
   &    O2 &      -3.01 &        18&&      -3.63 &        18 \\
   &    N3 &      -3.01 &        18&&      -3.94 &        17 \\
   &    N4 &      -3.35 &        22&&       -3.8 &        23 \\
   &    C8 &      -3.55 &        22&&       -3.8 &        23 \\
   &    Ht &       - &        -& &       - &        - \\
 \hline
\end{tabular}
\caption{Tabulated time (as) at which maximum flux (dQ(r,t)/dt) is observed at different atomic sites for keto and enol a" orbitals. No values are given when no substantial charge flux is present for the given atomic site.}
\label{tbl: flux a"}
\end{table*}

\end{document}